\documentclass{article}

\usepackage{cosmic-arxiv/arxiv}

\usepackage[utf8]{inputenc} % allow utf-8 input
\usepackage[T1]{fontenc}    % use 8-bit T1 fonts
\usepackage{hyperref} % hyperlinks
\usepackage{url} 
\usepackage{rotating}
\usepackage{algorithm}
\usepackage{algpseudocode}% simple URL typesetting
\usepackage{booktabs} 
\usepackage{comment} % professional-quality tables
\usepackage{amsfonts}       % blackboard math symbols
\usepackage{nicefrac}       % compact symbols for 1/2, etc.
\usepackage{microtype}  
\usepackage{amsmath} % microtypography
\usepackage{cleveref}       % smart cross-referencing
\usepackage{lipsum}         % Can be removed after putting your text content
\usepackage{graphicx}
\usepackage{natbib}
\usepackage{doi}
\usepackage{lscape}
\usepackage{multirow}
\usepackage{adjustbox}
\algnewcommand{\algorithmicforeach}{\textbf{for each}}
\algdef{SE}[FOR]{ForEach}{EndForEach}[1]
  {\algorithmicforeach\ #1\ \algorithmicdo}% \ForEach{#1}
  {\algorithmicend\ \algorithmicforeach}% \EndForEach
\title{Hierarchical Clustering in $\Lambda$CDM Cosmologies via Persistence Energy}

\newif\ifuniqueAffiliation
\uniqueAffiliationtrue

\ifuniqueAffiliation % Standard variant of author block
\author{Michael E.~Van Huffel\\
	Department of Mathematics\\
	ETH Zurich\\
	Zurich, Switzerland \\
	\texttt{michavan@student.ethz.ch} \\
	\And
	Leonardo A.~A. Barberi \\
	Department of Computer Science\\
	ETH Zurich\\
	Zurich, Switzerland \\
	\texttt{lbarberi@student.ethz.ch} \\
 	\And
	Tobias Sagis \\
	Department of Computer Science\\
	TU Eindhoven\\
	Eindhoven, Netherlands \\
	\texttt{t.g.m.sagis@student.tue.nl} \\
}
\else
\usepackage{authblk}

\newbox{\orcid}\sbox{\orcid}{\includegraphics[scale=0.06]{orcid.pdf}} 
\author[1]{%
	\href{https://orcid.org/0000-0000-0000-0000}{\usebox{\orcid}\hspace{1mm}David S.~Hippocampus\thanks{\texttt{hippo@cs.cranberry-lemon.edu}}}%
}
\author[1,2]{%
	\href{https://orcid.org/0000-0000-0000-0000}{\usebox{\orcid}\hspace{1mm}Elias D.~Striatum\thanks{\texttt{stariate@ee.mount-sheikh.edu}}}%
}
\affil[1]{Department of Computer Science, Cranberry-Lemon University, Pittsburgh, PA 15213}
\affil[2]{Department of Electrical Engineering, Mount-Sheikh University, Santa Narimana, Levand}
\fi

\renewcommand{\headeright}{}
\renewcommand{\undertitle}{}
\renewcommand{\shorttitle}{Hierarchical Clustering in $\Lambda$CDM Cosmologies via Persistence Energy}

\hypersetup{
pdftitle={A template for the arxiv style},
pdfsubject={q-bio.NC, q-bio.QM},
pdfauthor={David S.~Hippocampus, Elias D.~Striatum},
pdfkeywords={First keyword, Second keyword, More},
}

\begin{document}
\maketitle

\begin{abstract}
In this research, we investigate the structural evolution of the cosmic web, employing advanced methodologies from Topological Data Analysis. Our approach involves leveraging LITE, an innovative method from recent literature that embeds persistence diagrams into elements of vector spaces. Utilizing this methodology, we analyze three quintessential cosmic structures: clusters, filaments, and voids. A central discovery is the correlation between \textit{Persistence Energy} and redshift values, linking persistent homology with cosmic evolution and providing insights into the dynamics of cosmic structures.
\end{abstract}

% keywords can be removed

\section{Introduction}
\label{sec:intro}

Topological Data Analysis (TDA) has emerged as a transformative approach to extract meaningful information from complex datasets, offering a lens through which to understand the data's underlying structure. Unlike traditional data analysis methods that rely on geometric or statistical measures, TDA employs tools from both computational geometry and algebraic topology to study the topological features inherent in datasets. In the context of cosmology, where the distribution of matter exhibits complex and interconnected patterns, TDA becomes a valuable tool for uncovering the underlying cosmic topology. 

%\subsection{Problem Statement and Solution}

The cosmic web, encompassing galaxies, intergalactic gas, and dark matter, exhibits an organized tendency to form structures such as galaxy clusters, filaments (thread-like structures that connect galaxy clusters), and walls, surrounded by low-density void regions (\cite{Colberg2008, Weygaert2011, Cautun2014, Wilding_2021}). Within this cosmic context, large galaxy clusters aggregate into more extensive formations referred to as filaments or superclusters of galaxies \cite{Kelesis2022}.

% The scientific literature is replete with a plethora of computational methods proposed to enhance the understanding of the topological and geometrical patterns of the universe as a whole (refer to Section \ref{sec:relatedwork}). These computational methodologies affirm the intricate nature of the cosmic structure, constituting a challenging cosmic-scale network that eludes straightforward analytical approaches.

At the heart of our study is the understanding that the cosmic web is not a static entity. Its formation and evolution follow a hierarchical pattern: initially, small fluctuations in the density of dark matter lead to the creation of larger structures, primarily galaxy clusters. These clusters become gravitational anchors, pulling in nearby smaller galaxies. Over time, this process leads to the formation of even larger structures, as clusters merge and grow, continually reshaping the cosmic landscape \cite{Cautun2014}.

Simulations of the cosmic web are notoriously computationally expensive to perform analysis on. This is especially the case with standard TDA algorithms, such as the Vietoris-Rips filtration, that are not particularly well-designed for computational efficiency. In response to the computational challenges posed by the cosmic network, we present a specialized TDA pipeline crafted to address the unique demands of cosmic web analysis. 

Our study focuses on the dynamic nature of the cosmic web, which evolves hierarchically from small dark matter density fluctuations to larger structures like galaxy clusters. These clusters, acting as gravitational centers, attract smaller galaxies and merge to form even larger cosmic structures. To analyze this complex and computationally intensive cosmic web, we introduce a specialized TDA pipeline, designed for efficiency and tailored to the unique requirements of cosmic web analysis.

In the first step of our pipeline, we summarize cosmic information using the well-established Delaunay Tessellation Field Estimator (DTFE) method (\cite{Schaap2000, Weygaert2008, Kelesis2022, Wilding_2021}). 
%Widely applied in the examination of the cosmic web and its evolutionary dynamics, this method entails reconstructing a volumetric representation using continuous density and intensity fields derived from a discrete set of spatial points.
To effectively capture and analyze the intricate features of the cosmic web, a filtration process based on cubical complexes is applied on the DTFE output. 
%The adoption of cubical complexes is driven by their proficiency in dealing with data over rectangular domains, such as images or meshes, as exemplified by the density values derived from the mesh output of the DTFE method \cite{Sousbie2011}.
%This choice significantly enhances the interpretability and computational efficiency of the subsequent analysis, making cubical complexes particularly well-suited for navigating the complexities inherent in the cosmic web.

%The complex geometry inherent the construction of persistence diagrams presents significant challenges, particularly due to the high dimensionality of comic datasets, which leads to an excessive number of birth-death points. This complexity hampers the direct applicability and interpretability of these diagrams, complicating the extraction of meaningful insights. To address this issue, we incorporate the \textit{Spectral Persistence Homology} framework into our pipeline. This innovative approach reinterprets persistence diagrams, which naturally arise as discrete measures in \(\mathbb{R}^2_+\), treating them as signals. These signals are then embedded into vector spaces using functional-based transforms via \textit{Persistence Signals} (PSs) vectorization method, \cite{vanhuffel2023spectral}.  After embedding the diagrams, we analyze the what we call \textit{persistence energy} of PSs, observing changes over different redshift values spanning \(z = 5\) to \(z = 0\).

The complex geometry of persistence diagrams presents significant interpretive challenges for extracting meaningful insights. To address this, our methodology integrates a framework for embedding these diagrams into a vector space. This approach is based on the recent work of \cite{vanhuffel2023spectral} and further explained in Section \ref{sec:van}. Subsequently, we analyze what we term the \textit{persistence energy} of these diagrams across varying redshift values, ranging from $z = 5$ to $z = 0$. 

The structure of this paper is organized as follows: Section \ref{sec:preliminaries:cosmicweb} provides foundational knowledge about the cosmic web, including the relevant terminology used throughout this paper. Section \ref{sec:data} details the dataset and sources employed in our research. In Section \ref{sec:relatedwork}, we discuss related work and our contributions to the field. Section \ref{sec:methodology} outlines our methodology, explaining the processes and techniques used in our study. Section \ref{sec:results} presents our results, offering insights and implications drawn from our analysis. Finally, Section \ref{sec:conclusion} concludes the paper with a summary of our findings and potential directions for future research. Additionally, the implementation of our work is made available as an open-source resource\footnote{\href{https://github.com/majkevh/cosmic-master}{https://github.com/majkevh/cosmic-master}}.
\subsection{The Cosmic Web}
\label{sec:preliminaries:cosmicweb}
The cosmic web is a large-scale structure that delineates the distribution of matter in the observable universe. It is a vast network of interconnected filaments, voids, and clusters that span billions of light-years. At its core, the cosmic web provides a structural framework for understanding the arrangement of galaxies and cosmic matter on the largest scales. Key components of the cosmic web include:

\begin{itemize}
    \item \textit{Clusters}: Dense regions where numerous galaxies and dark matter concentrations coalesce.
    \item \textit{Filaments}: Long, thread-like structures that connect galaxy clusters.
    \item \textit{Voids}: Expansive, nearly empty regions separating filaments and clusters \cite{Wilding_2021}.
\end{itemize}

In this work, we present an analysis of the evolution of the structure of the cosmic web, based on simulations within the $\Lambda$CDM cosmological model. Furthermore, we analyse multiple snapshots at different moments in time, as captured by the redshift value.

\textit{Snapshots} of the cosmic web capture a moment in the evolution of the large-scale structure of the universe. In the context of simulations or observational surveys, a snapshot represents a specific instance in time, typically corresponding to a particular redshift value. The \textit{redshift} value associated with a snapshot is a measure of the cosmic expansion; it quantifies the extent to which the universe has stretched since the emitted light from distant objects was initially produced \cite{lambdacdm}. Higher redshift values correspond to earlier cosmic times, allowing researchers to study the cosmic web at different epochs. Essentially, a snapshot with a higher redshift captures the universe when it was younger, providing insights into the formation and evolution of structures such as filaments, voids, and galaxy clusters over cosmic time scales. 

One central thesis regarding the cosmic web is that the components that characterize the cosmic web, namely cluster, filaments, and voids, merge together, such that the present day web is dominated by fewer, but much more massive, structures \cite{lambdacdm}.

\section{Data Collection and Sources}\label{sec:data}

In our study, we employ the TNG50-1-Dark dataset from the IllustrisTNG simulations, which model the universe's evolution following a $\Lambda$CDM cosmology, to analyze the distribution and dynamics of dark matter. 

IllustrisTNG is a set of cosmological hydrodynamical simulations that model the evolution of the universe, including the formation and evolution of galaxies, dark matter, and other cosmic structures \cite{nelson2021illustristng}. It consists of three distinct simulation volumes: TNG50, TNG100, and TNG300. These volumes are tailored to specific scientific objectives, varying in physical size, mass resolution, and the complexity of included physics. The cubic volumes, with side lengths of approximately 50, 100, and 300 Mpc for TNG50, TNG100, and TNG300 respectively, serve as the canvas for the simulations.

%Each simulation volume operates at three resolution levels. The highest resolution simulations employ over 20, 10, and 30 billion resolution elements for TNG50, TNG100, and TNG300 respectively. Importantly, the physical model is deliberately constructed to be resolution-independent, maintaining consistency across different resolution levels. The simulations assume a $\Lambda \mathrm{CDM}$ cosmological model with parameters calibrated to the Planck constraints (Planck Collaboration 2015). Important computed parameters in this list include but are not limited to matter density $\Omega_m=0.3089\pm0.0062$, dark energy density $\Omega_{\Lambda}=0.6911\pm0.0062$ and Hubble's constant $H_0=67.74\pm0.46 \ \mathrm{km} \ s^{-1}\mathrm{Mpc}^{-1}$

In this study, we focus on  multiple redshifts to study the chronological evolution and strengthen the results obtained through our proposed methods using the TNG50-1-Dark ("dark matter only") periodic volume, analyzing structural formations (\cite{nelson2019tng50}, \cite{pillepich2019tng50}). Dark-matter-only simulations give predictions for how the large scale structure, the clustering of galaxies, the shapes of halos, and so forth would evolve in a universe constructed only of dark matter. 

%In the TNG50-1-Dark dataset, we obtain information related to the distribution, dynamics, and properties of dark matter within the simulated cosmological volume. This includes data on dark matter halos, which are concentrations of dark matter that serve as the gravitational seeds for galaxy formation, as well as details on the large-scale structure of dark matter filaments and voids.

For our study, we consider only the 3-dimensional comoving coordinates of dark matter subhalos, measured in \( ckpc/h \), spanning from redshift \( z = 5 \) to \( z = 0 \). This treats the dataset as a point cloud. Due to machine limitations, we use only a cropped cube of the dataset with a side length of approximately \( 2.5 \times 10^4 \) ckpc/h.

\section{Related Work and Contribution}\label{sec:relatedwork}

Recent advances in the quality and quantity of observational data from surveys pertaining to $\Lambda$CDM simulations has led to great progress in understanding the topological structure and connectivity of the cosmic web. A large number of computational methods have been proposed, with the objective to further our understanding of the topological patterns of the universe, confirming the structure of a web-like network, while also raising curiosity on the intrinsic properties of the cosmic network, namely connectivity and complexity \cite{Kelesis2022}.
%Several computational methods have also been proposed to investigate the spatial connectedness of galaxies (\cite{coutinho2016network, Klypin1983}). In \cite{coutinho2016network}, for example, the authors explore several network construction algorithms that use various galaxy properties to assign a network to galaxy distributions, presenting a novel approach in the use of graph theory to visualise and understand the cosmic web.

The application of TDA methodologies to the cosmic web, while theoretically sensible, is still a rather novel approach in the analysis of the cosmos. In \cite{Wilding_2021}, the authors present a persistent homology pipeline in their study of the cosmic web, following a similar structure to what is done in this study. In their work, they introduce interpretations on the evolution of the cosmic web by parametrizing Betti curves and quantitatively analysing the evolution of the parameters that characterise their distribution.

%Finally, in the context of analysis of persistence diagrams, \cite{cao2023kmeans} proved the convergence of k-means clustering algorithms on this highly complex space, establishing strong theoretical properties for this type of analysis, which will be presented in Section \ref{sec:PCA_results}. 

Our research introduces the use of a new vectorization framework of TDA in analyzing the cosmic web, validating it in a novel context. This application underscores the hierarchical clustering of cosmic structures, reinforcing the $\Lambda$CDM universe model. We study the evolution  of the \textit{persistence energy} of PSs across different redshifts, offering insights into the cosmic web's structural dynamics.

\section{Background}

Persistent homology is a cornerstone technique in TDA, renowned for its effectiveness in analyzing complex data structures such as point clouds, time-series, and graphs. This method operates by tracking homological features across varying scales. It utilizes a filtration process to construct a nested sequence of topological spaces (\( X_0 \subseteq X_1 \subseteq \cdots \subseteq X_n = X \)), enabling a comprehensive topological exploration. 

At the heart of this approach are persistence diagrams, which are crucial for visualizing and understanding the data's topological features. These diagrams plot points in the half-plane $\Omega \cup \partial \Omega$, where $\Omega=\{(x, y) \in \mathbb{R}^2 | x<y\}$ and $\partial \Omega=\{(x, x) \in \mathbb{R}^2\}$. Each point in a persistence diagram represents a topological feature, with its \textit{birth} (when it appears) and \textit{death} (when it disappears) during the filtration process. The coordinates of a point \((x, y)\) indicate the scale at which the feature appears and vanishes, respectively. The distance of a point from the diagonal. signifies the \textit{persistence} or longevity of the feature, with points farther from the diagonal indicating features that persist over a wider range of scales. 

Persistence diagrams are characteristic for topological features in distinct dimensions. $p$-dimensional topological features are captured by the {homology group} $H_p$, and the collection of all diagrams from $p$-dimensional homology is denoted by \( \mathcal{D}_p \). For those new to this field, further insights into these concepts can be gained from foundational texts such as \cite{edelsbrunner2017persistent} and \cite{DeyWang2022}.

In their work, \cite{chazal2013structure} reinterprets a persistence diagram as a measure \( \mu = \sum_{\mathbf x \in D} m(\mathbf x) \delta_{\mathbf x} \), where \( D \subset \Omega \) is locally finite, and each \( \mathbf x \) has an associated multiplicity \( m(\mathbf x) \in \mathbb{N} \) (notation from \cite{vanhuffel2023spectral}). Building upon this, \cite{divol2019understanding} introduces \( \mathcal{M}^p \), the space of Radon measures with support on \(\Omega\) and a finite \( p \) value, also referred to as the \textit{persistence measures space}. By integrating \( \mathcal{M}^p \) with the optimal partial transport distance, as introduced by \cite{figalli2010new}, a metric space \( (\mathcal{M}^p, \mathrm{OT}_p) \) is defined.\footnote{The specifics of $OT_p$ are not detailed here as they are not crucial for this work.} Unlike \( \mathcal{D}_p \), the space \( \mathcal{M}^p \) provides a more suitable framework for statistical analyses. This is because persistence measures are linear objects, simplifying the definition and computation of statistical quantities like means, as discussed by \cite{divol2021estimation} and \cite{cao2022approximating}.

\label{sec:preliminaries}

\section{Methodology}
\label{sec:methodology}

\begin{figure}[ht]
\centering
\includegraphics[width=.9\textwidth]{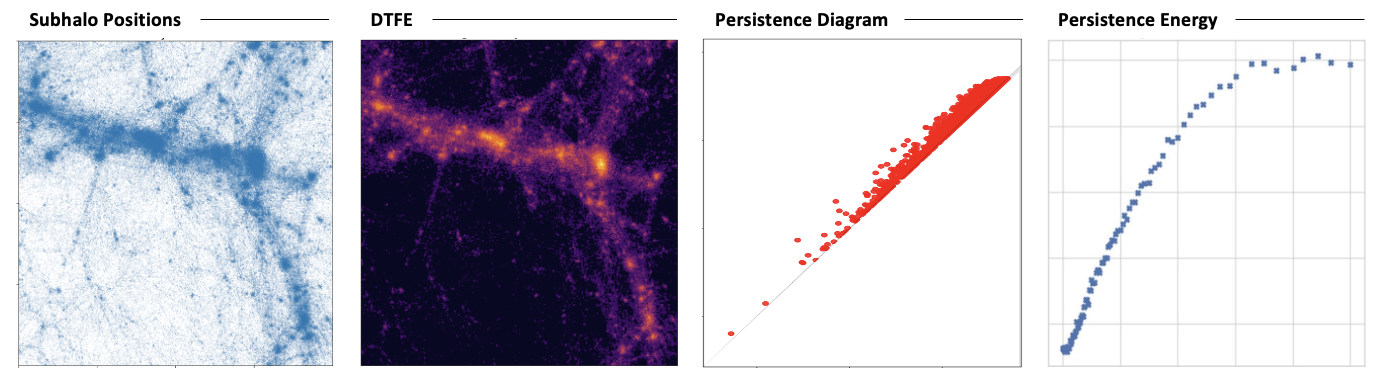}
\caption{Overarching methodology presented in this paper.}
\label{fig:research_diagram}
\end{figure}

The methodology presented in this paper is illustrated in Figure \ref{fig:research_diagram}. In this pipeline, the initial input is a 3D point cloud representing dark matter subhalos. This point cloud undergoes processing through the DTFE algorithm. As described by \cite{Schaap2000}, the DTFE algorithm interpolates each point within the cloud onto a fixed-size grid, assigning a density value to each grid point. Following this, a super-level-set filtration of these density values is conducted using cubical complexes, a process efficiently detailed in the works of \cite{wagner2011efficient} and \cite{ mischaikow2004computational}.  Once the persistence diagram of this filtration is constructed, it is then vectorized employing methods such as LITE described in \cite{vanhuffel2023spectral}, and \textit{persistence energy} (see Section \ref{sec:energy}).

\subsection{DTFE}\label{sec:DTFE}
Originally proposed by Schaap W.E., van de Weygaert R, the Delaunay Tessellation Field Estimator is a mathematical tool used for reconstructing density fields from a discrete set of samples/measurements \cite{Schaap2000,van_de_Weygaert_2008,cautun2019dtfe}. Using the maximum information contained within the point set, such as velocity statistics, DTFE can construct very precise, high resolution continuous density and intensity fields. 

The method is based on the stochastic geometric concept of the Delaunay tessellation generated by the point set. Given a set of $N$ discrete points, in a finite region in $\mathbb{R}^d$, the Delaunay tessellation $\mathcal{D}$ is constructed. The size of the triangles formed is a measure of the local density of the point distribution. This property is exploited in the second step of the algorithm, by estimating the local density at the sampling points using linear interpolation. In the final step of the algorithm, the obtained density estimates are interpolated to any other point \cite{Schaap2000}.

The DTFE method is highly suitable for astrophysical data. The local geometry of the point distribution is preserved, which allows for finding simplices of varying dimensions, including filaments, walls and voids \cite{Wilding_2021}. An additional advantage of this method is that it does not depend on user-defined choices or parameters.

\begin{figure}
    \centering
    \includegraphics[width=.6\textwidth]{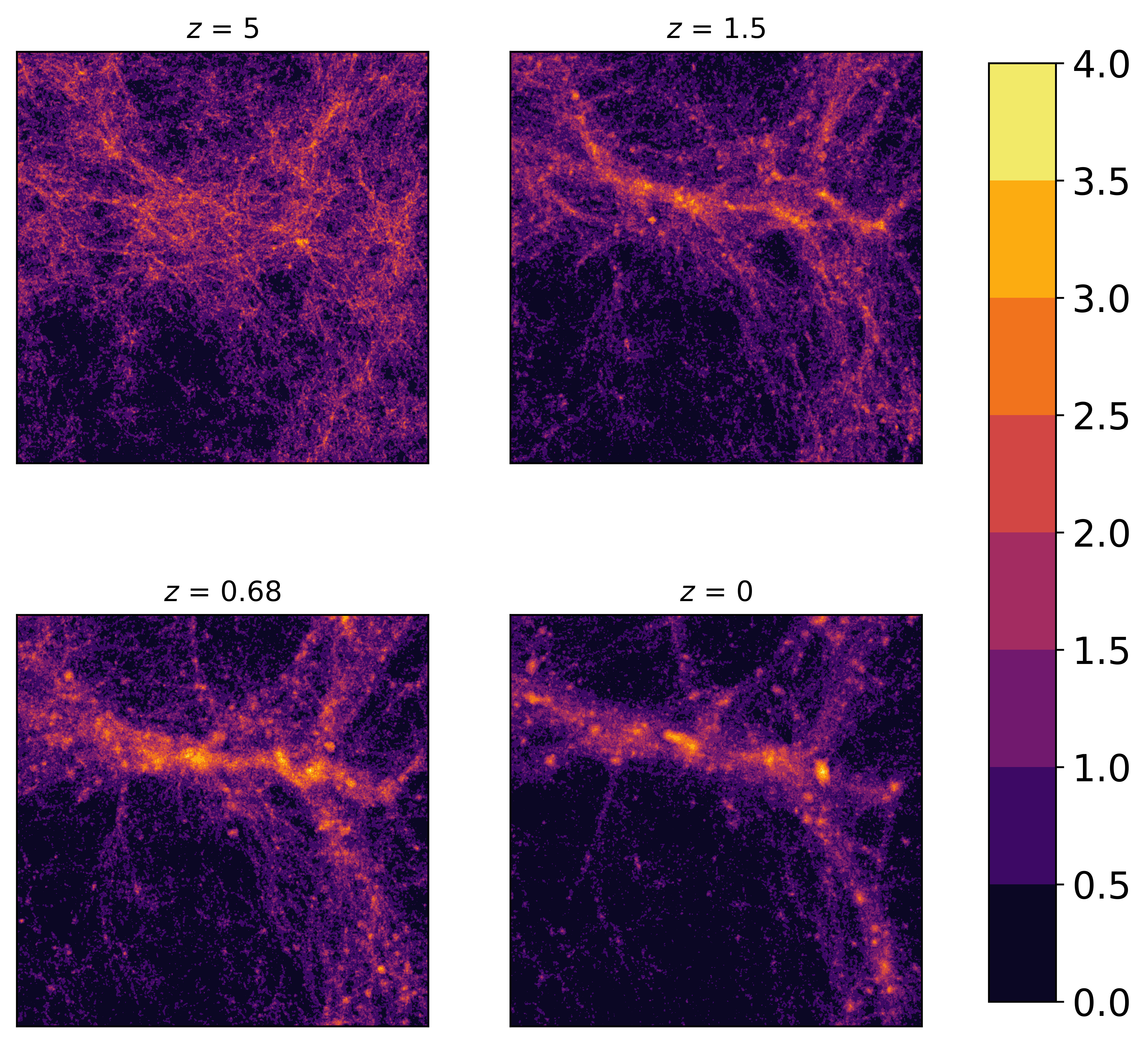}
    \caption{DTFE log-density estimation at various redshifts.}
    \label{fig:dtfe_redshifts}
\end{figure}

As explained in Section \ref{sec:data}, our dataset consists of partitioned sets of dark matter subhalos' moving coordinates. Each sub-partition is considered as a set of $N$ discrete points in a finite region in $\mathbb{R}^3$, thus serving as primary input for the DTFE algorithm. Finally, a grid size $g$ determines the cardinality of the output matrix $M$, of dimensions $(g \times g \times g)$. At each grid index $(i,j,k) \in M$, the corresponding interpolated density estimate can be found. The output of the DTFE from the original subhalo positions at different redshifts can be seen in Figure  \ref{fig:dtfe_redshifts}.

\subsection{Filtration of subhalo galaxies positions}
%In our research, we begin by examining the positions of sub-halo dark matter particles. To address the computational and conceptual complexities inherent in studying the cosmic network, we first implement the DTFE on these particle positions. The DTFE, a technique extensively used in studying the cosmic web and its evolutionary dynamics, allows us to reconstruct a volumetric representation. It does this by transforming a discrete set of spatial points into continuous density and intensity fields, offering a more manageable and informative view of the cosmic structure.

Following the DTFE application, we proceed to analyze the intricate features of the cosmic web through a filtration process using cubical complexes, a process efficiently detailed in the works of \cite{wagner2011efficient} and \cite{ mischaikow2004computational}. Cubical complexes are particularly adept at handling data in rectangular domains, such as images or meshes, making them a fitting choice for working with the density values obtained from the DTFE mesh output. This approach is underpinned in \cite{Sousbie2011}, which highlights the effectiveness of cubical complexes in cosmic web analysis. By employing cubical complexes, we significantly enhance the interpretability and computational efficiency of our analysis. 

The filtration process we adopt is a superlevel set filtration based on the density values derived from the DTFE output. This method allows us to systematically analyze the cosmic web's structure at various density levels, providing a comprehensive understanding of its topological features.

\subsection{Embedding of Persistence Diagrams}\label{sec:vectorisation}
Vectorizing persistence diagrams is essential for interpreting topological findings, as it provides a structured and quantitative way to represent complex topological data. Direct interpretation of persistence diagrams can be difficult due to their intricate nature and the challenges in identifying subtle patterns. Although qualitative analysis is occasionally feasible, it becomes particularly challenging with persistence diagrams that contain numerous topological features and \textit{noise} (points near the diagonal), a situation common when dealing with data arising from enourmous datasets like the ones in our study. As highlighted in \cite{Bubenik} and \cite{coniglio}, merely removing this noise is insufficient, as these features often play a crucial role in describing the underlying shape of the data. 

\subsubsection{Lattice Integrated Topological Embedding - (LITE)}\label{sec:van}

LITE represent an innovative category of descriptors within TDA, selected for their computational efficiency, theoretical grounding, and simplicity, \cite{vanhuffel2023spectral}. This approach  first shift persitence diagrams into a \textit{birth-persistence} coordinate system and then transform them  into a elements of a finite-dimensional vector space using a functional transform. For an in-depth explanation of this transformation process, we refer readers to \cite{vanhuffel2023spectral}. In this work, we focus on two types of functional transforms: the straightforward identity transform and the more complex Gabor functional transform.

The Gabor functional is particularly adept at providing a comprehensive view of the topological features in the diagrams. Its ability to offer both a local and global perspective stems from its nature as a windowed Fourier transform. This means it applies a localized Fourier transform to different segments of the persistence diagram. As a result, it captures detailed, localized information (the local perspective) while also preserving the broader, overall structure of the data (the global perspective). Consequently, the Gabor functional is especially suitable for TDA, where understanding both the fine details and the larger picture of the data's topology is crucial.

\subsubsection{Persistence Energy}\label{sec:energy}
Persistence energy is a metric derived from the transformation coefficients of LITE. This concept is inspired by the classical notion of Fourier energy in signal processing. Analogously, the \textit{persistence energy} of a measure $\mu\in\mathcal M^p$ can be mathematically expressed as:
\begin{equation}\label{eq:energy} E_{\Psi_f}(\mu) = \sum_{\mathbf x\in \Gamma_p} |\kappa_f(\mathbf x)|^2\,, \end{equation}
where \( \kappa(\mathbf x) \) denotes the transformation coefficient, which depends on the choice of function $f$ as specified in Equation (\ref{eq:energy}). Furthermore, \(\Gamma_p\) is defined as the space of discretized persistence diagrams (DPDs), a measure subspace of $\mathcal M^p$ in which measures are confined to allocating mass exclusively at points on a pre-defined  grid. For more comprehensive details on this space, readers are referred to \cite{vanhuffel2023spectral}. 

The \textit{persistence energy} quantifies the significance of topological features represented in a persistence diagram for a measure \(\nu \in \Gamma_p\). This metric provides a scalar summary, enabling easier comparison and analysis of topological signatures. 

In this work, owing to the substantial heterogeneity in the multiplicity of different birth-persistence pairings in the DPDs, we precede the computation of the \textit{persistence energy} of the  measure $\nu$ with the application of a logarithmic transform on the DPDs. This transformation is essential for normalizing the varying scales of the birth-persistence pairings, thereby facilitating a more uniform and accurate analysis.

\begin{comment}
    
Persistence energy, inspired by Fourier energy in signal processing, is a metric from the transformation coefficients of PSs. Mathematically, it's expressed as:
\begin{equation}
E_{\Psi_f}(\mu) = \sum_{\mathbf x\in \Gamma_p} |\kappa_f(\mathbf x)|^2\,,
\end{equation}
where \( \kappa(\mathbf x) \) is the transformation coefficient dependent on function $f$, and \(\Gamma_p\) represents the space of discretized persistence diagrams (DPDs), \cite{vanhuffel2023spectral}. This metric quantifies the significance of topological features in a persistence diagram, providing a scalar summary for comparison and analysis. Due to the heterogeneity in DPDs' birth-persistence pairings, we precede the computation of the persistence energy with the application of a logarithmic transform to normalize these scales, ensuring a uniform and accurate analysis.
\end{comment}

\section{Results}
\label{sec:results}

\begin{figure}[H]
    \centering
    \includegraphics[width=\textwidth]{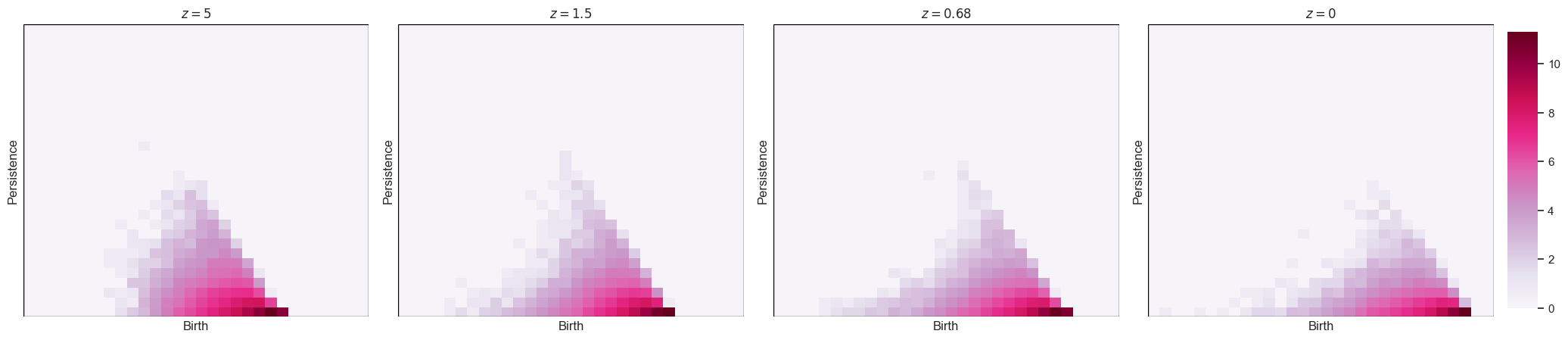}
    \caption{DPDs of $H_{1}$ at different redshifts. Heatmap scale is in logarithm.}
    \label{fig:H1DPD}
\end{figure}

%Figure \ref{fig:H1DPD} presents the evolution of DPDs for the \(H_1\) homology group. These diagrams are derived from the cubical complex superlevelset filtration applied to the DTFE output of the cosmic web. The evolution is shown across various redshift values \(z\). For comparison, similar plots for the \(H_0\) and \(H_2\) homology groups can be found in Figure \ref{fig:HODPD} and Figure \ref{fig:H2DPD}, respectively. It's important to remember that a higher redshift value signifies a greater distance from the present in temporal terms, with redshift \(z=0\) representing the most recent snapshot of the cosmic web. 
Figure \ref{fig:H1DPD} shows the evolution of DPDs for the \(H_1\) homology group, derived from the cubical complex superlevelset filtration on the DTFE output of the cosmic web, across various redshift values \(z\). For context, higher redshift values indicate earlier cosmic times, with \(z=0\) being the most recent. Comparable plots for \(H_0\) and \(H_2\) homology groups are in Figure \ref{fig:HODPD} and Figure \ref{fig:H2DPD}, respectively.

\begin{figure}[h]
    \centering
    \includegraphics[width=0.9\textwidth]{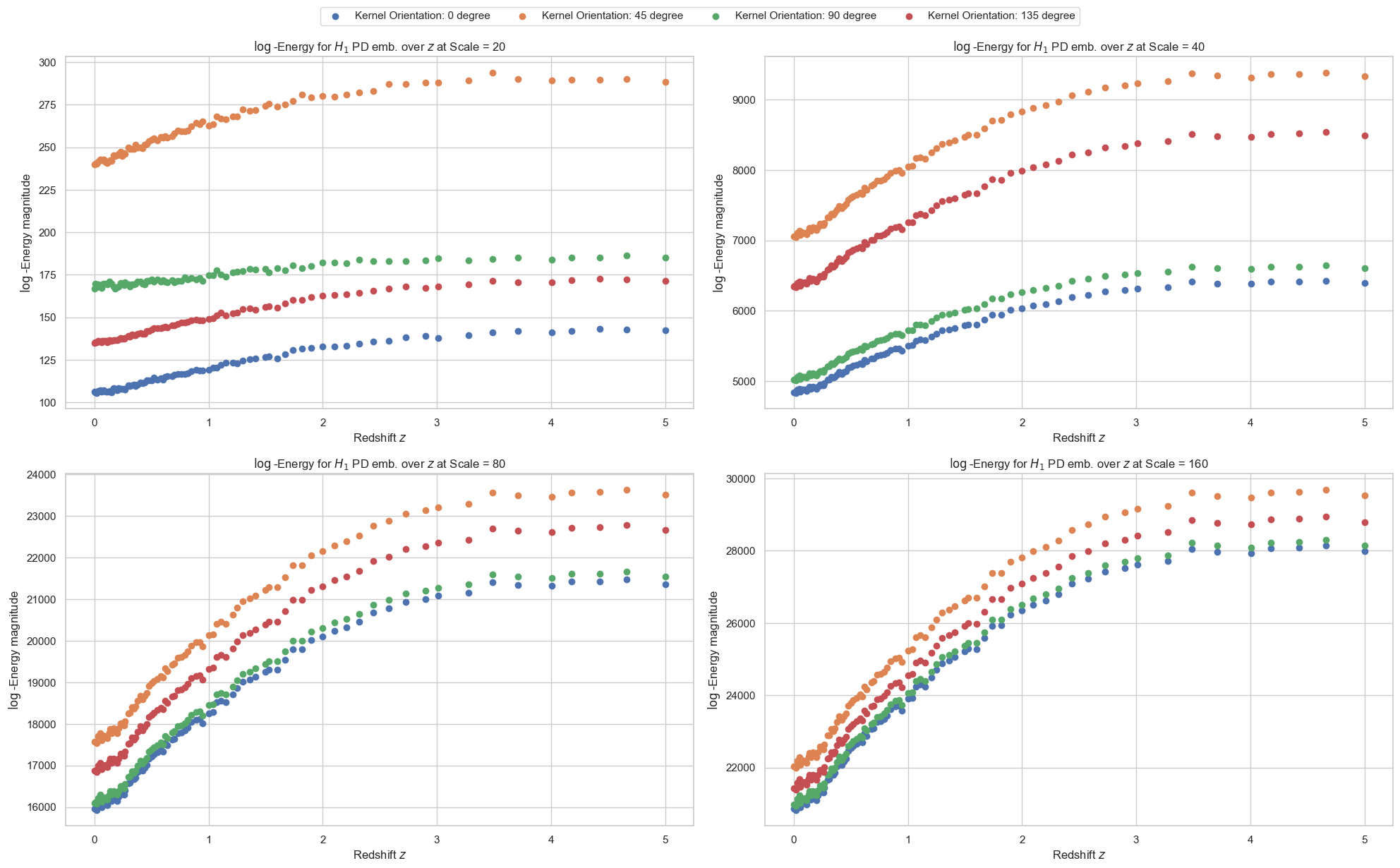}
    \caption{Persistence energy variation with redshifts and Gabor parameters for $H_1$ diagram.}
    \label{fig:GABBO2}
\end{figure}

A qualitative analysis of these plots is sufficient to discern meaningful patterns. Across all homology groups, we observe that, as the redshift value decreases (approaching present time), there is a notable shift in the topological activity towards the right along the birth-axis. Furthermore, as supported in Figure \ref{fig:GABBO2}, the overall energy of the diagrams decrease globally and locally, with the phenomenon being more prominent in the global case.

The decrease in energy indicates that, over time, there are increasingly fewer loops created by inter-galactic filaments. The rightward shift in Figure \ref{fig:H1DPD} indicates that the loops are formed by filaments containing an increasing density of galaxies. We observe the same phenomenon for 0- and 2-dimensional features in Figures \ref{fig:HODPD} and \ref{fig:H2DPD}, further reinforcing the notion that the cosmic web undergoes a process of merging over time, with each characteristic structure being defined on higher density levels. Similar results are found in \cite{Wilding_2021} by analysing persistence curves, a well-established vectorization of persistence diagrams.

Moreover, the DPDs of all three homology groups exhibit a high density of points near zero persistence, consistent across all redshift values.  This observation is particularly relevant when examining the energy changes at different scales using Gabor transform of the DPDs (refer to Figure \ref{fig:GABBO1}, \ref{fig:GABBO2}, \ref{fig:GABBO3}). At lower scales, which are sensitive to finer details, the energy values show minimal variation over $z$. This stability suggests that these fine-grained features, likely representing minor fluctuations (noise) or micro-scale patterns, remain relatively unchanged throughout the observed epochs. Since these features do not change significantly over different redshifts, it reinforces the idea that they might be inconsequential fluctuations rather than meaningful structures.

In contrast, at higher scales, there is a marked and consistent decrease in energy, indicating a substantial alteration in the larger-scale features. This drastic reduction in energy at higher scales can be attributed to the diminishing prominence of significant, high-persistence features in the cosmic web. Such features, which might represent more substantial structural formations or dense regions, appear to reduce in number or prominence as the universe evolves towards its current state, confirming the thesis that galaxies tend to merge together in an increasingly dense cosmic web \cite{Cautun2014, lambdacdm}. These observations, when considered together, further reinforce the $\Lambda$CDM  model of the universe. 

\begin{figure}
    \centering
    \includegraphics[width=.9\textwidth]{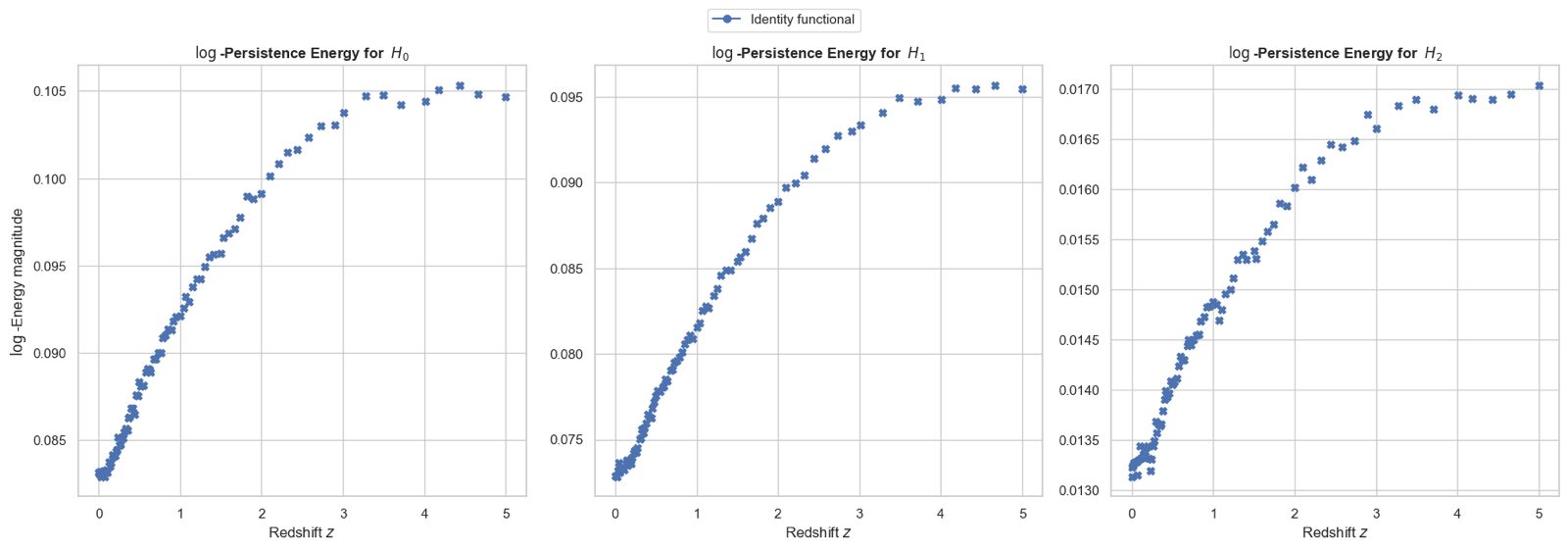}
    \caption{log-Persistence energy for the first three homology groups, using the identity function as a transform.}
    \label{fig:identity-energy}
\end{figure}

Figure \ref{fig:identity-energy} displays the log of the \textit{persistence energy} for the first three homology groups, using the identity function as functional. As with the Gabor transforms, we observe a similar trend in the \textit{persistence energy} as the cosmic web evolves. This is a testament to the robustness and efficiency of the methodology presented in this paper.

%These observations, put together, further support the thesis that the galaxy clusters in the cosmic web tend to merge together and form more dense isolated structures, filaments and voids \cite{lambdacdm}.

\section{Conclusions and further work}
\label{sec:conclusion}

\subsection{Conclusions}
Our research utilized TDA to investigate the evolving structure of the cosmic web. We conducted a thorough examination of the development of cosmic clusters, filaments, and voids across a range of redshift values. The findings from this study offer robust support for the $\Lambda$CDM model. Specifically, the observed trend in \textit{persistence energy} at different scales, as revealed through the application of the Gabor transform, provides critical insights into the hierarchical clustering and merging processes that are fundamental to the formation of the universe's large-scale structure. This analysis not only validates existing cosmological theories but also underscores the significance of topological methods in deepening our understanding of the universe's evolution.

\subsection{Further work}
In our future work, we aim to refine and enhance our methodologies in several key areas. Firstly, we plan to implement a non-regular or adaptive grid for Delaunay Triangulation Field Estimation (DTFE). The use of cubical complexes on regular grids has led to anisotropy issues and a challenge in capturing fine details in areas where they are most needed. A more adaptive approach could potentially address these concerns.

Moreover, we are considering an alternate approach as proposed by \cite{Anthea2022} in  \textit{Approximating Persistent Homology for Large Datasets}. This novel method involves using a measure-based approach to approximate persistence homology for large datasets through bootstrap techniques. It focuses on computing the Mean Persistence Measure and Fréchet means of Persistence Diagrams. However, the stability of this method has not yet been proven, indicating an area ripe for further exploration and validation.

These directions not only aim to improve the alignment of our computational methods with theoretical models but also strive to enhance the precision and reliability of our analysis in understanding the complex nature of the cosmic web.

\newpage
\bibliographystyle{unsrtnat}
\bibliography{cosmic-arxiv/references} 

\begin{thebibliography}{27}
\providecommand{\natexlab}[1]{#1}
\providecommand{\url}[1]{\texttt{#1}}
\expandafter\ifx\csname urlstyle\endcsname\relax
  \providecommand{\doi}[1]{doi: #1}\else
  \providecommand{\doi}{doi: \begingroup \urlstyle{rm}\Url}\fi

\bibitem[{Colberg} et~al.(2008){Colberg}, {Pearce}, {Foster}, {Platen}, {Brunino}, {Neyrinck}, {Basilakos}, {Fairall}, {Feldman}, {Gottl{\"o}ber}, {Hahn}, {Hoyle}, {M{\"u}ller}, {Nelson}, {Plionis}, {Porciani}, {Shandarin}, {Vogeley}, and {van de Weygaert}]{Colberg2008}
J.~M. {Colberg}, F.~{Pearce}, C.~{Foster}, E.~{Platen}, R.~{Brunino}, M.~{Neyrinck}, S.~{Basilakos}, A.~{Fairall}, H.~{Feldman}, S.~{Gottl{\"o}ber}, O.~{Hahn}, F.~{Hoyle}, V.~{M{\"u}ller}, L.~{Nelson}, M.~{Plionis}, C.~{Porciani}, S.~{Shandarin}, M.~S. {Vogeley}, and R.~{van de Weygaert}.
\newblock \textit{The Aspen-Amsterdam void finder comparison project}.
\newblock \emph{\textup{mnras}}, 387\penalty0 (2):\penalty0 933--944, June 2008.
\newblock \doi{10.1111/j.1365-2966.2008.13307.x}.

\bibitem[Van~de Weygaert and Platen(2011)]{Weygaert2011}
R.~Van~de Weygaert and E.~Platen.
\newblock \textit{COSMIC VOIDS: STRUCTURE, DYNAMICS AND GALAXIES}.
\newblock \emph{\textup{International Journal of Modern Physics: Conference Series}}, 01:\penalty0 41--66, 2011.
\newblock \doi{10.1142/S2010194511000092}.
\newblock URL \url{https://doi.org/10.1142/S2010194511000092}.

\bibitem[Cautun et~al.(2014)Cautun, van~de Weygaert, Jones, and Frenk]{Cautun2014}
M.~Cautun, R.~van~de Weygaert, B.~J.~T. Jones, and C.~S. Frenk.
\newblock \textit{Evolution of the cosmic web}.
\newblock \emph{\textup{Monthly Notices of the Royal Astronomical Society}}, 441\penalty0 (4):\penalty0 2923--2973, 05 2014.
\newblock ISSN 0035-8711.
\newblock \doi{10.1093/mnras/stu768}.
\newblock URL \url{https://doi.org/10.1093/mnras/stu768}.

\bibitem[Wilding et~al.(2021)Wilding, Nevenzeel, van~de Weygaert, Vegter, Pranav, Jones, Efstathiou, and Feldbrugge]{Wilding_2021}
G.~Wilding, K.~Nevenzeel, R.~van~de Weygaert, G.~Vegter, P.~Pranav, B.~J.~T. Jones, K.~Efstathiou, and J.~Feldbrugge.
\newblock \textit{Persistent homology of the cosmic web – I. Hierarchical topology in $\Lambda$CDM cosmologies}.
\newblock \emph{\textup{Monthly Notices of the Royal Astronomical Society}}, 507\penalty0 (2):\penalty0 2968–2990, August 2021.
\newblock ISSN 1365-2966.
\newblock \doi{10.1093/mnras/stab2326}.
\newblock URL \url{http://dx.doi.org/10.1093/mnras/stab2326}.

\bibitem[Kelesis et~al.(2022)Kelesis, Basilakos, Papadopoulou, Fotakis, and Efstathiou]{Kelesis2022}
D.~Kelesis, S.~Basilakos, V.~Papadopoulou, L., D.~Fotakis, and A.~Efstathiou.
\newblock \textit{Detecting and analysing the topology of the cosmic web with spatial clustering algorithms I: methods}.
\newblock \emph{\textup{Monthly Notices of the Royal Astronomical Society}}, 516\penalty0 (4):\penalty0 5110–5124, September 2022.
\newblock ISSN 1365-2966.
\newblock \doi{10.1093/mnras/stac2444}.
\newblock URL \url{http://dx.doi.org/10.1093/mnras/stac2444}.

\bibitem[{Schaap} and {van de Weygaert}(2000)]{Schaap2000}
W.~E. {Schaap} and R.~{van de Weygaert}.
\newblock \textit{Continuous fields and discrete samples: reconstruction through Delaunay tessellations}.
\newblock \emph{\textup{aap}}, 363:\penalty0 L29--L32, November 2000.
\newblock \doi{10.48550/arXiv.astro-ph/0011007}.

\bibitem[{van de Weygaert} and {Bond}(2008)]{Weygaert2008}
R.~{van de Weygaert} and J.~R. {Bond}.
\newblock {Clusters and the Theory of the Cosmic Web}.
\newblock In M.~{Plionis}, O.~{L{\'o}pez-Cruz}, and D.~{Hughes}, editors, \emph{A Pan-Chromatic View of Clusters of Galaxies and the Large-Scale Structure}, volume 740, page 335. 2008.
\newblock \doi{10.1007/978-1-4020-6941-3_10}.

\bibitem[Van~Huffel and Palo(2024)]{vanhuffel2023spectral}
M.~E. Van~Huffel and M.~Palo.
\newblock \textit{LITE: A Stable Framework for Lattice-Integrated Embedding of Topological Descriptors}, 2024.

\bibitem[{Planck Collaboration}(2020)]{lambdacdm}
{Planck Collaboration}.
\newblock \textit{Planck 2018 results - VI. Cosmological parameters}.
\newblock \emph{\textup{AA}}, 641:\penalty0 A6, 2020.
\newblock \doi{10.1051/0004-6361/201833910}.
\newblock URL \url{https://doi.org/10.1051/0004-6361/201833910}.

\bibitem[Nelson et~al.(2021)Nelson, Springel, Pillepich, Rodriguez-Gomez, Torrey, Genel, Vogelsberger, Pakmor, Marinacci, Weinberger, Kelley, Lovell, Diemer, and Hernquist]{nelson2021illustristng}
D.~Nelson, V.~Springel, A.~Pillepich, V.~Rodriguez-Gomez, P.~Torrey, S.~Genel, M.~Vogelsberger, R.~Pakmor, F.~Marinacci, R.~Weinberger, L.~Kelley, M.~Lovell, B.~Diemer, and L.~Hernquist.
\newblock \textit{The IllustrisTNG Simulations: Public Data Release}, 2021.

\bibitem[Nelson et~al.(2019)Nelson, Pillepich, Springel, Pakmor, Weinberger, Genel, Torrey, Vogelsberger, Marinacci, and Hernquist]{nelson2019tng50}
D.~Nelson, A.~Pillepich, V.~Springel, R.~Pakmor, R.~Weinberger, S.~Genel, P.~Torrey, M.~Vogelsberger, F.~Marinacci, and L.~Hernquist.
\newblock \textit{First results from the TNG50 simulation: galactic outflows driven by supernovae and black hole feedback}.
\newblock \emph{\textup{Monthly Notices of the Royal Astronomical Society}}, 490\penalty0 (3):\penalty0 3234–3261, August 2019.
\newblock ISSN 1365-2966.
\newblock \doi{10.1093/mnras/stz2306}.
\newblock URL \url{http://dx.doi.org/10.1093/mnras/stz2306}.

\bibitem[Pillepich et~al.(2019)Pillepich, Nelson, Springel, Pakmor, Torrey, Weinberger, Vogelsberger, Marinacci, Genel, van~der Wel, and Hernquist]{pillepich2019tng50}
A.~Pillepich, D.~Nelson, V.~Springel, R.~Pakmor, P.~Torrey, R.~Weinberger, M.~Vogelsberger, F.~Marinacci, S.~Genel, A.~van~der Wel, and L.~Hernquist.
\newblock \textit{First results from the TNG50 simulation: the evolution of stellar and gaseous discs across cosmic time}.
\newblock \emph{\textup{Monthly Notices of the Royal Astronomical Society}}, 490\penalty0 (3):\penalty0 3196–3233, September 2019.
\newblock ISSN 1365-2966.
\newblock \doi{10.1093/mnras/stz2338}.
\newblock URL \url{http://dx.doi.org/10.1093/mnras/stz2338}.

\bibitem[Edelsbrunner and Morozov(2017)]{edelsbrunner2017persistent}
H.~Edelsbrunner and D.~Morozov.
\newblock \emph{Persistent Homology}.
\newblock CRC Press, 3 edition, 2017.
\newblock To appear.

\bibitem[Dey and Wang(2022)]{DeyWang2022}
T.~K. Dey and Y.~Wang.
\newblock \emph{Computational Topology for Data Analysis}.
\newblock Cambridge University Press, 2022.
\newblock ISBN 9781009098168.

\bibitem[Chazal et~al.(2013)Chazal, de~Silva, Glisse, and Oudot]{chazal2013structure}
F.~Chazal, V.~de~Silva, M.~Glisse, and S.~Oudot.
\newblock The structure and stability of persistence modules.
\newblock \emph{arXiv preprint arXiv:1207.3674}, Mar 2013.

\bibitem[Divol and Lacombe(2019)]{divol2019understanding}
V.~Divol and T.~Lacombe.
\newblock Understanding the topology and the geometry of the space of persistence diagrams via optimal partial transport.
\newblock \emph{arXiv preprint arXiv:1901.03048}, 2019.

\bibitem[Figalli and Gigli(2010)]{figalli2010new}
A.~Figalli and N.~Gigli.
\newblock A new transportation distance between non-negative measures, with applications to gradients flows with dirichlet boundary conditions.
\newblock \emph{Journal de Math{\'e}matiques Pures et Appliqu{\'e}es}, 94\penalty0 (2):\penalty0 107--130, 2010.

\bibitem[Divol and Lacombe(2021)]{divol2021estimation}
V.~Divol and T.~Lacombe.
\newblock Estimation and quantization of expected persistence diagrams, 2021.

\bibitem[Monod(2022)]{cao2022approximating}
Yueqi Monod.
\newblock Approximating persistent homology for large datasets.
\newblock \emph{arXiv preprint arXiv:2204.09155}, 2022.

\bibitem[Wagner et~al.(2011)Wagner, Chen, and Vu{\c{c}}ini]{wagner2011efficient}
H.~Wagner, C.~Chen, and E.~Vu{\c{c}}ini.
\newblock \textit{Efficient computation of persistent homology for cubical data}.
\newblock In \emph{\textup{Topological methods in data analysis and visualization II: theory, algorithms, and applications}}, pages 91--106. Springer, 2011.

\bibitem[Mischaikow et~al.(2004)Mischaikow, Kaczynski, and Mrozek]{mischaikow2004computational}
K~Mischaikow, T~Kaczynski, and M~Mrozek.
\newblock \textit{Computational homology}.
\newblock \emph{\textup{Applied Mathematical Sciences}}, 157, 2004.

\bibitem[van~de Weygaert and Schaap(2008)]{van_de_Weygaert_2008}
R.~van~de Weygaert and W.~Schaap.
\newblock \emph{The Cosmic Web: Geometric Analysis}, page 291–413.
\newblock Springer Berlin Heidelberg, 2008.
\newblock ISBN 9783540447672.
\newblock \doi{10.1007/978-3-540-44767-2_11}.
\newblock URL \url{http://dx.doi.org/10.1007/978-3-540-44767-2_11}.

\bibitem[Cautun and van~de Weygaert(2019)]{cautun2019dtfe}
M.~C. Cautun and R.~van~de Weygaert.
\newblock The dtfe public software: The delaunay tessellation field estimator code, 2019.

\bibitem[Sousbie(2011)]{Sousbie2011}
T.~Sousbie.
\newblock The persistent cosmic web and its filamentary structure - i. theory and implementation: Persistent cosmic web - i: Theory and implementation.
\newblock \emph{Monthly Notices of the Royal Astronomical Society}, 414\penalty0 (1):\penalty0 350–383, April 2011.
\newblock ISSN 0035-8711.
\newblock \doi{10.1111/j.1365-2966.2011.18394.x}.
\newblock URL \url{http://dx.doi.org/10.1111/j.1365-2966.2011.18394.x}.

\bibitem[Bubenik(2020)]{Bubenik}
P.~Bubenik.
\newblock \emph{The Persistence Landscape and Some of Its Properties}, pages 97--117.
\newblock 06 2020.
\newblock ISBN 978-3-030-43407-6.
\newblock \doi{10.1007/978-3-030-43408-3_4}.

\bibitem[Hacquard(2023)]{coniglio}
O.~Hacquard.
\newblock \textit{Statistical learning on measures: an application to persistence diagrams}.
\newblock \emph{\textup{arXiv preprint arXiv:2303.08456}}, 2023.

\bibitem[Cao and Monod(2022)]{Anthea2022}
Y.~Cao and A.~Monod.
\newblock Approximating persistent homology for large datasets, 2022.

\end{thebibliography}

\newpage
\appendix
%\lipsum[1-2]
\section{Appendix}
In this section we present the results of the Vectorization Techniques used in this work. We present DPD's of homology groups $H_0, H_1$ and $H_2$, as well as the \textit{persistence energy variation} diagrams for the aforementioned homology groups.

\begin{figure}[h]
    \centering
    \includegraphics[width=\textwidth]{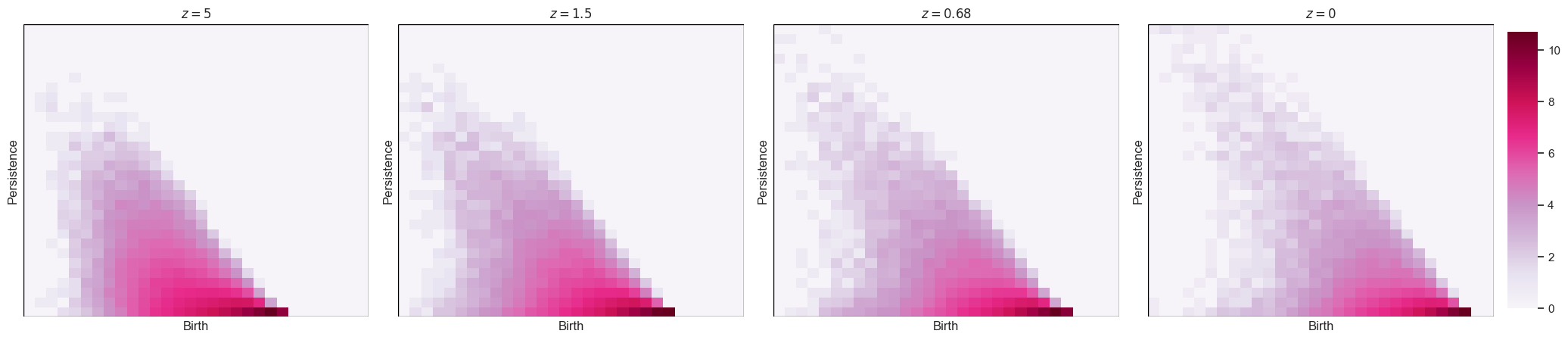}
    \caption{DPDs of $H_{0}$ at different redshifts. Heatmap scale is in logarithm.}
    \label{fig:HODPD}
\end{figure}
\begin{figure}[h]
    \centering
    \includegraphics[width=\textwidth]{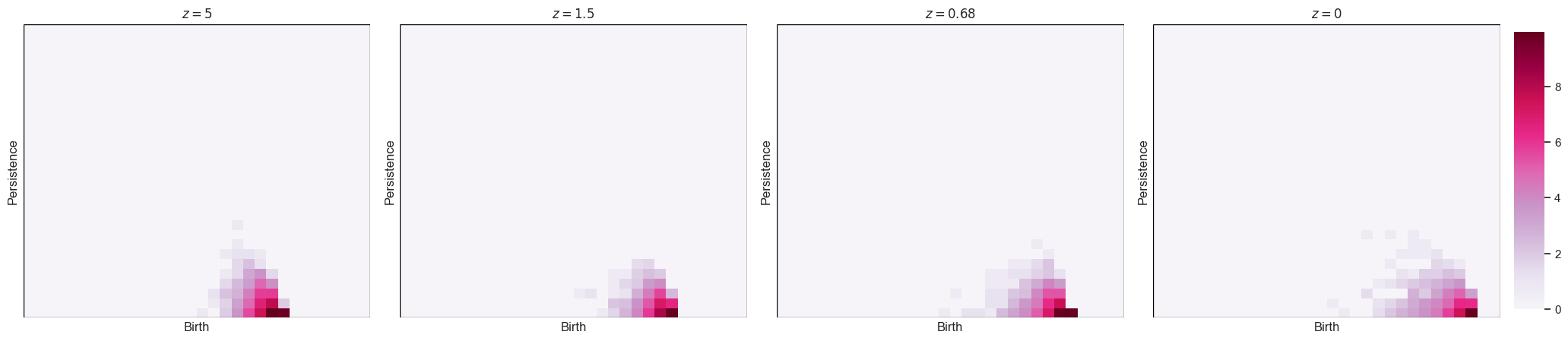}
    \caption{DPDs of $H_{2}$ at different redshifts. Heatmap scale is in logarithm.}
    \label{fig:H2DPD}
\end{figure}
\begin{figure}[h]
    \centering
    \includegraphics[width=0.9\textwidth]{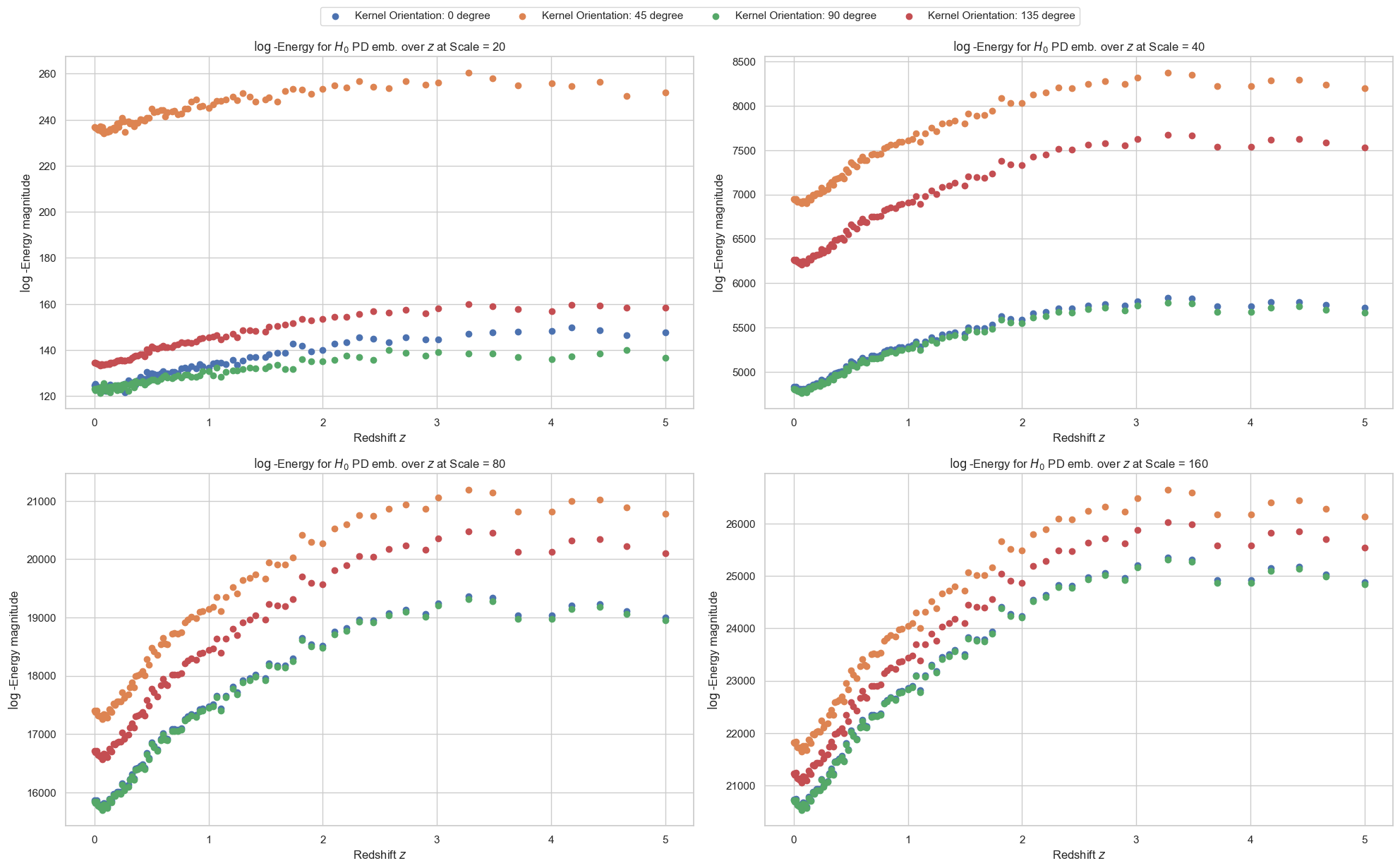}
    \caption{Persistence energy variation with redshifts and Gabor parameters for $H_0$ diagram.}
    \label{fig:GABBO1}
\end{figure}

\begin{figure}[h]
    \centering
    \includegraphics[width=0.9\textwidth]{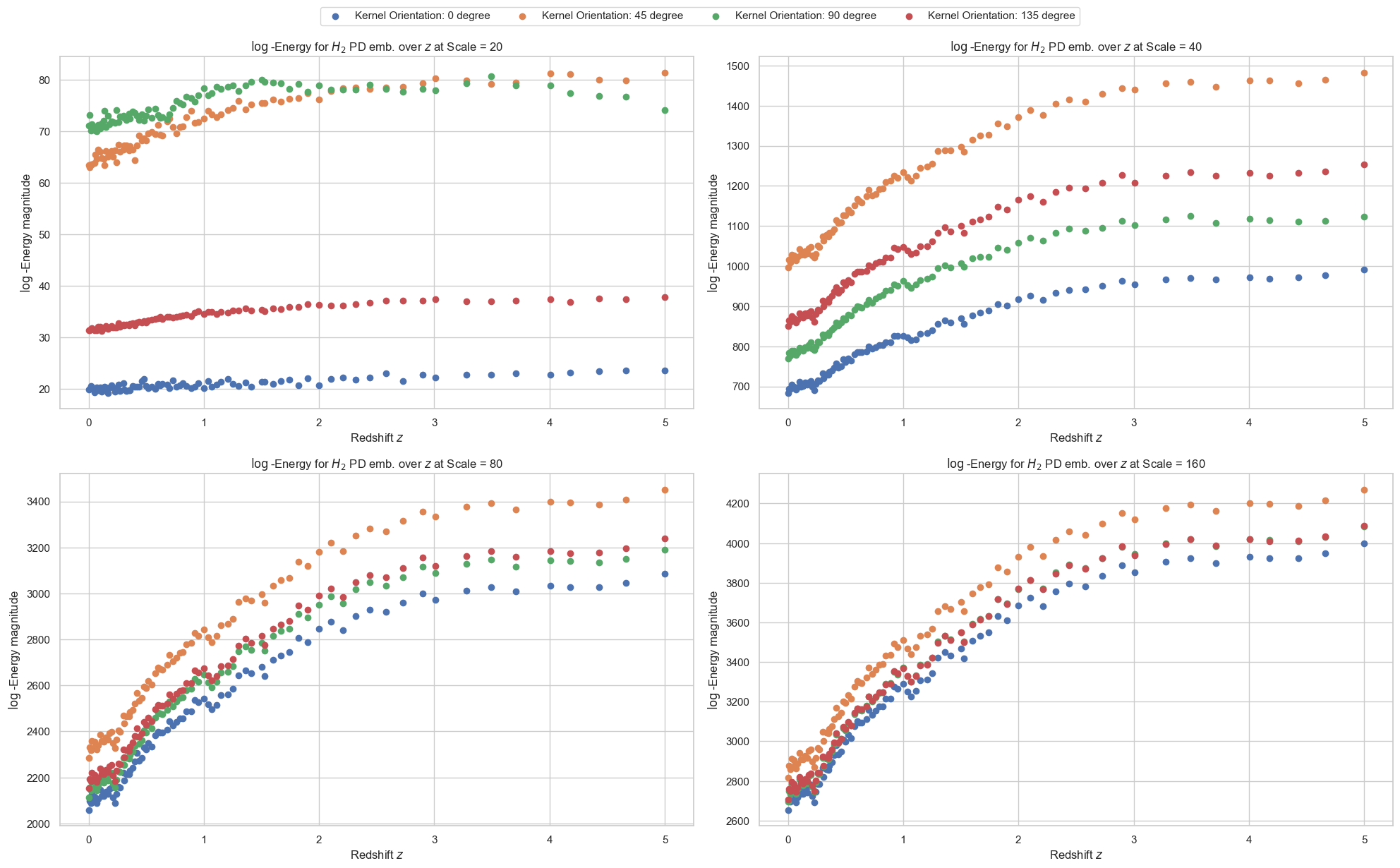}
    \caption{Persistence energy variation with redshifts and Gabor parameters for $H_2$ diagram.}
    \label{fig:GABBO3}
\end{figure}

\end{document}